\documentclass[reprint,eqsecnum,prb]{revtex4-1}

\usepackage{graphicx}
\usepackage{amsmath}
\usepackage{color}
\usepackage{subfigure}

\begin{document}

\title{On the theoretical description of weakly charged surfaces}

\author{Rui Wang}

\author{Zhen-Gang Wang}
\email {zgw@caltech.edu}
\affiliation{Division of Chemistry and Chemical Engineering,California Institute of Technology, Pasadena, CA 91125, USA}


\begin{abstract}

It is widely accepted that the Poisson-Boltzmann (PB) theory
provides a valid description for charged surfaces in the so-called
weak coupling limit. Here, we show that the image charge repulsion
creates a depletion boundary layer that cannot be captured by a
regular perturbation approach.  The correct weak-coupling theory
must include the self-energy of the ion due to the image charge
interaction. The image force qualitatively alters the double layer
structure and properties, and gives rise to many non-PB effects,
such as nonmonotonic dependence of the surface energy on
concentration and charge inversion.  In the presence of dielectric
discontinuity, there is no limiting condition for which the PB
theory is valid.

\end{abstract}

 \maketitle

\section{Introduction}
\label{sec:level1}

The electric double layer resulting from a charged surface in an
aqueous solution affects a wealth of structural and dynamic
properties in a wide range of physicochemical, colloidal,
soft-matter and biophysical
systems\cite{Israelachvili,Gelbart,Levinreview,Honig,Luo}. The
standard textbook description of the electrical double layers is
based on the mean-field Poisson-Boltzmann (PB) theory. At large
surface-charge density, high counter-ion valency and high ion
concentration -- the so-called strong coupling limit -- it is well
recognized that PB theory fails to capture a number of qualitative
effects, such as like-charge attraction\cite{Kjellander,Bowen,Jho}
and charge
inversion\cite{Kekicheff,Shklovskii1,Shklovskii2,Besteman}.
Liquid-state theories\cite{Kjellander2,Monica} and other
strong-coupling theories\cite{Jho,Netz} have been employed to
account for the strong ion-ion correlations in this regime.

In the opposite limit -- the weak-coupling regime -- it is generally
accepted that the electric double layer is well described by the PB
theory\cite{Netz,Netzandorland,Ninham1,Ninham2,Podgornik,Podgornik2,Netz2,Neu,Kanduc,Podgornikrev}.
Performing a loop-wise perturbation expansion\cite{Netzandorland} in
the coupling parameter (to be defined below), Netz\cite{Netz2}
demonstrated that the PB theory is the leading-order theory in the
weak-coupling limit, and becomes exact in the limit of zero coupling
strength. Applying Netz's approach explicitly to surfaces with
dielectric discontinuity, Kandu$\check{\rm c}$ and
Podgornik\cite{Kanduc} concluded that, under the weak-coupling
condition, the image force only enters as a small correction to the
leading PB theory, which vanishes in the limit of zero coupling.  In
particular, the self-energy due to image charge interaction was
shown not to appear in the Boltzmann factor for the ion
distributions.  Although these demonstrations were performed
explicitly for counterion-only systems, the conclusions are
generally believed to hold when salt ions are
added\cite{Podgornikrev}.   Thus, many researchers in the
electrolyte community consider the weak-coupling theory to mean the
PB theory; in other words, weak coupling is considered synonymous
with the validity of the PB theory.

Physically, however, a single ion in solution next to a surface of a
lower dielectric plate obviously should feel the image charge
repulsion even in the absence of any surface charge, and the ion
distribution -- the probability of finding the ion at any location
-- should reflect the image charge interaction through the Boltzmann
factor.  This was the case studied in the pioneering work of
Wagner\cite{wagner}, and Onsager and Samaras\cite{onsager} (WOS) for
the surface tension of electrolyte solutions.  It is rather odd that
this interaction should become absent from the Boltzmann factor for
the distribution of mobile ions in the weak-coupling limit when the
surface becomes charged.  It is also rather curious that the image
interaction, which is absent from the Boltzmann factor in the
Netz-Kandu$\check{\rm c}$-Podgornik (NKP)
approach\cite{Netz,Netz2,Kanduc} in the weak coupling limit,
``re-emerges" in the Boltzmann factor in the strong-coupling limit,
though in a different form (through a fugacity
expansion)\cite{Jho,Kanduc,Podgornikrev}. Taking zero-surface charge
as the limiting case of the {\it physical} weak-coupling condition,
it is clear that the NKP and WOS approaches give drastically
different descriptions of the same system. It is also difficult to
physically reconcile the absence of the image interaction from the
Boltzmann factor in the weak-coupling limit with its ``re-emergence"
in the strong coupling limit in the NKP approach. Furthermore, ion
depletion near a weakly-charged dielectric interface has been
observed in Monte Carlo simulation\cite{Netz,Levinsimulation} as
well as predicted by the hypernetted chain approximation (HNC)
integral equation theory that includes the image charge
interactions\cite{HNC}.


In this work, we clarify the origin of these discrepancies by a
re-examination of the role of the image charge interaction in the
{\it physical} weak-coupling limit.  We show that in the presence of
a dielectric discontinuity, the {\it physical} weak-coupling limit
is not described by the so-called weak-coupling theory if the latter
is meant to be the PB or PB with small fluctuations corrections. The
image charge repulsion creates a boundary layer which cannot be
captured by the the NKP approach.
 A nonperturbative approach yields a modified Poisson-Boltzmann equation, where a screened,
 self-consistently determined image charge interaction appears in the Boltzmann factor for the ion concentration
for any surface charge density.  The WOS theory is an approximation
of the more general framework presented here in the special case of
zero surface charge.

To see the origin of the boundary layer, we start by an analysis of
the relevant length scales for the counterion-only system. Consider
a charged planar surface at $z=0$ with charge density $\sigma$
separating an aqueous solution ($z>0$) from an semi-infinite plate
($z<0$). The solvent and plate are taken to be dielectric continuum
with dielectric constant $\varepsilon_S$ and $\varepsilon_P$,
respectively, with $\varepsilon_P<<\varepsilon_S$.   Now consider a counterion of valency $q$ at distance
$z$ away from the surface.  The attraction between the test ion and
the charged surface is $E_{sur}=2\pi q l_B \sigma z = z/l_{GC}$,
whereas the repulsion due to its image charge is $E_{im}=f q^2 l_B
/(2z)$, where $l_B=e^2/(4\pi \varepsilon_0\varepsilon_S kT)$ is the
Bjerrum length with $\varepsilon_0$ denoting the vacuum
permittivity, $l_{GC}=1/(2 \pi q \sigma l_{B})$ is the Gouy-Chapman
length and
$f=(\varepsilon_S-\varepsilon_P)/(\varepsilon_S+\varepsilon_P)$
represents the dielectric contrast between the two media.  Balancing
$E_{sur}$ with $E_{im}$ results in a characteristic length:
\begin{equation}
d=\left(f/2\right)^{1/2}q \left(l_B l_{GC}\right)^{1/2}
\label{eq1.1}
\end{equation}
Introducing the coupling parameter $\Xi=q^2 l_B/l_{GC}$,\cite{Netz2}
we see
\begin{equation}
d \sim l_B \Xi^{-1/2} \ \ and \ \ d/l_{GC} \sim \Xi^{1/2}
\label{eq1.2}
\end{equation}
Thus, as the coupling strength $\Xi$ goes to zero, $d$
itself diverges, but the ratio of $d$ to $l_{GC}$ (noting that
$l_{GC}$ is the characteristic length scale for the double layer in
the PB theory) goes to zero. This is a typical feature of a {\it
boundary layer}. Physically, the competition between the surface
charge attraction and the image charge repulsion gives rise to a
depletion boundary layer.  Since the perturbation approach performs
an expansion in powers of $\Xi$\cite{Netz,Netz2,Kanduc} (which
results from nondimensionalizing all the lengths by the longest
length scale $l_{GC}$), information within the smaller length-scale
-- the depletion boundary layer -- is lost.  Although this analysis
is performed explicitly for the counterion-only system, the
depletion boundary layer persists when salt ions are introduced.

\section{A Gaussian Variational Approach}
\label{sec:level2}

The presence of a boundary layer necessitates a nonperturbative
treatment. Using the renormalized Gaussian variational
approach\cite{Orland}, one of us\cite{Wang1} derived a general
theory for electrolyte solutions with dielectric
inhomogeneity. In this section, we first recapitulate the key steps
in the derivation of the general theory and then specify to the case
of a charged plate with dielectric discontinuity.

\subsection{General Theory}
\label{sec:levelA}

We consider a general system with a fixed charge distribution $e
\rho_{ex}({\bf r})$ in the presence of small mobile cations of
valency $q_+$ and anions of valency $q_+$ in a dielectric medium of
a spatially varying dielectric function $\varepsilon ({\bf r})$. $e$
is the elementary charge. The charge on the ion is assumed to have a
finite spread given by a short-range distribution function $h_{\pm}
({\bf r}-{\bf r}_i)$ for the {\it i}th ion, with the point-charge
model corresponding to $h_{\pm} ({\bf r}-{\bf r}_i)=q_{\pm} \delta
({\bf r}-{\bf r}_i)$. The introduction of a finite charge
distribution on the ion avoids the divergence of the short-range
component of the self energy -- the local solvation energy --
resulting from the point-charge model, and reproduces the Born
solvation energy\cite{Wang1}. However, as the emphasis of this work
is on the long-range component of the self energy -- the image
charge interaction -- which is finite for point charges, we will
eventually take the point-charge limit for the ion. The diverging
but constant local solvation energy in the point-charge limit can be
regularized by subtracting the same-point Green function in the
bulk, as discussed below.  Since we work in the low concentration
regime for the ions ($c \le 0.1M$) (the Debye-H\"uckel regime), the
excluded volume effects of the ions are unimportant, and so we treat
the ions as volumeless.

The total charge density including both the fixed charges and mobile ions is
\begin{eqnarray}
e \rho({\bf r}) = &e& \rho_{ex}({\bf r}) +e \int d{\bf r}' [
h_+ ({\bf r}'-{\bf r})\hat{c}_+({\bf r}')  \nonumber \\
&-&h_- ({\bf r}'-{\bf r})\hat{c}_-({\bf r}')   ] \label{eq2.1}
\end{eqnarray}
with $ \hat{c}_{\pm} ({\bf r}) = \sum_{i=1}^{n_{\pm}} \delta ({\bf
r}-{\bf r}_i) $ the particle density operator for the ions. The
Coulomb energy of the system, {\it including the self energy}, is
\begin{equation}
H= \frac{e^2}{2 } \int d{\bf r} d {\bf r}' \rho({\bf r})G_0 ({\bf
r},{\bf r}' )\rho({\bf r}') \label{eq2.2}
\end{equation}
where $G_0 ({\bf r},{\bf r}' )$ is the Coulomb operator given by
\begin{equation}
- \nabla \cdot \left[\varepsilon_0 \varepsilon ({\bf r}) \nabla
G_0({\bf r},{\bf r}' ) \right] = \delta({\bf r}-{\bf r}' )
\label{eq2.3}
\end{equation}
It is convenient to work with the grand canonical partition function
\begin{eqnarray}
\Omega=\sum_{n_+=0}^{\infty} \sum_{n_-=0}^{\infty} \frac{{\rm e}
^{n_+ \mu_+}  {\rm e}^{ n_-  \mu_-}}{n_+!n_-! v_+^{n_+} v_-^{n_-}} \nonumber \\
\times \int \prod_{i=1}^{n_+} d {\bf r}_i \prod_{j=1}^{n_-} d {\bf
r}_j \exp \left(- \beta H \right) \label{eq2.4}
\end{eqnarray}
where $\mu_{\pm}$ are the chemical potential
for the cations and anions, and $v_{\pm}$ are some characteristic volume scales, which have no
thermodynamic consequence. We perform the usual
Hubbard-Stratonovich transformation to Eq. \ref{eq2.4} by
introducing a field variable $\phi({\bf r})$, which yields
\begin{equation}
\Omega= \frac{1}{Z_0}  \int D \phi \exp \left\{ - L [\phi] \right\}
\label{eq2.5}
\end{equation}
The ``action" $L$ is
\begin{eqnarray}
L=   \int d {\bf r} &[& \frac{1}{2}  \epsilon (\nabla \phi)^2 + i
\rho_{ex} \phi - \Gamma \lambda_+ {\rm e}^{- i  {\hat h}_+ \phi }\nonumber \\
&-&  \Gamma \lambda_-  {\rm e}^{ i  {\hat h}_-\phi } ] \label{eq2.6}
\end{eqnarray}
$Z_0$ is the normalization factor given by
\begin{equation}
Z_0= \int D \phi \exp \left[- \frac{1}{2}  \int d {\bf r} \epsilon
(\nabla \phi)^2  \right] = \left[\det  {\bf G_0}\right]^{1/2}
\label{eq2.7}
\end{equation}
where $G_0^{-1}= \nabla_{{\bf r}} \cdot \left[ \epsilon ({\bf r})
\nabla_{{\bf r}'} \delta ({\bf r}-{\bf r}') \right]$ is the inverse of
the Coulomb operator in Eq. \ref{eq2.3},  $\epsilon =\varepsilon_0
\varepsilon/(\beta e^2)$ is a scaled permittivity, and
$\lambda_{\pm}={\rm e}^{ \mu_{\pm}}/ v_{\pm}$ is the fugacity of the
ions. We have used the short-hand notation ${\hat h}_{\pm} \phi$ to
represent the local spatial averaging of $\phi$ by the charge
distribution function: ${\hat h}_{\pm} \phi= \int d{\bf r}' h_{\pm}
({\bf r}'-{\bf r}) \phi ({\bf r}')$. The function $\Gamma({\bf r})$
in Eq. \ref{eq2.6} is introduced to constrain the mobile ions to the
solvent region.

Equations \ref{eq2.5} and \ref{eq2.6} are the exact field-theoretic
representation for the partition function. Because the action is
nonlinear, the partition function cannot be evaluated exactly.  The
lowest-order approximation corresponds to taking the saddle-point
contribution of the functional integral, which results in the
Poisson-Boltzmann equation.  A systematic loop expansion can be
developed to account for fluctuations around the saddle point in an
order by order manner.  In practice, most theoretical treatments
only include one-loop corrections.  The loop expansion involves
expanding the action around the saddle point in polynomial forms.
However, the fluctuation part of the electrostatic potential due to
image charge interaction becomes very large near the dielectric
interface; thus any finite-order expansion of the $\langle {\rm
e}^{\mp i {\hat h}_{\pm} \phi } \rangle$ term, which becomes the
Boltzmann factor in the ion distribution (see Eq. \ref{eq2.11}), is
problematic.  The absence of the image-charge self-energy in the
Boltzmann factor in the perturbation
approaches\cite{Ninham1,Ninham2,Netz2,Kanduc,Podgornik,Podgornik2,Podgornikrev}
is thus a consequence of low-order expansion of the exponential
function of an imaginary variable when the variable can be quite
large.

To develop a nonperturbative theory, we perform a variational
calculation of Eq. \ref{eq2.5} using the Gibbs-Feynman-Bogoliubov
bound for the grand free energy $W = -\ln \Omega$, which yields
\begin{equation}
W  \le -\ln \Omega_{ref} +  \langle L[\phi]- L_{ref}[\phi] \rangle
\label{eq2.8}
\end{equation}
where
\begin{equation}
\Omega_{ref}=\frac{1}{Z_0} \int D \phi \exp \left\{
-L_{ref}\left[\phi \right]\right\} \label{eq2.9}
\end{equation}
The average $\langle \cdots \rangle$ is taken in the reference
ensemble with the action $L_{ref}$. We take the reference action to be
of the Gaussian form centered around the mean $-{i \psi}$
\begin{equation}
L_{ref}=\frac{1}{2} \int d {\bf r} d {\bf r}' [\phi ({\bf r})+ {i
\psi} ({\bf r})] G^{-1} ({\bf r},{\bf r}') [\phi ({\bf r}') +{i
\psi}({\bf r}')] \label{eq2.10}
\end{equation}
where $G^{-1}$ is the functional inverse of the Green function
$G$, and the introduction of $i$ is to keep
the mean electrostatic potential $\psi$ real. $\psi$ and $G$ are taken to be variational parameters for the
grand free energy functional.

With the Gaussian reference action Eq. \ref{eq2.10}, all the terms
on the r.h.s. of Eq. \ref{eq2.8} can be evaluated analytically (see
Appendix A for detailed derivation). The lower bound of the free
energy is obtained by extremizing the r.h.s. of Eq. \ref{eq2.8} with
respect to $\psi$ and $G$, which results in the following two
variational conditions:
\begin{equation}
- \nabla \cdot \left( \epsilon \nabla \psi \right) = \rho_{ex} +
\Gamma  \lambda_+  q_+ {\rm e}^{ - q_+  \psi -  u_+ } - \Gamma
\lambda_- q_- {\rm e}^{q_-  \psi - u_- } \label{eq2.11}
\end{equation}
\begin{equation}
- \nabla \cdot \left[ \epsilon \nabla G({\bf r},{\bf r}') \right] +
2  I({\bf r}) G({\bf r},{\bf r}')=  \delta ({\bf r}-{\bf r}')
\label{eq2.12}
\end{equation}
where $u_{\pm}$ is the self energy of the ions
\begin{equation}
u_{\pm} ({\bf r})= \frac{1}{2}   \int d {\bf r}'  d {\bf r}''
h_{\pm} ({\bf r}-{\bf r}') G({\bf r}',{\bf r}'') h_{\pm} ({\bf
r}''-{\bf r}) \label{eq2.13}
\end{equation}
$I({\bf r})=\left[q_+^2 c_+({\bf r})+q_-^2 c_-({\bf r})\right]/2$ is
the local ionic strength, with the concentration of cations and
anions given by
\begin{equation}
c_{\pm} ({\bf r}) =\lambda_{\pm} \Gamma \exp \left[ \mp q_{\pm} \psi
({\bf r}) -u_{\pm} ({\bf r}) \right] \label{eq2.14}
\end{equation}

Eqs. \ref{eq2.11}-\ref{eq2.13} forms a set of self-consistent
equations for the mean electrostatic potential $\psi({\bf r})$, the
correlation function (Green function) $G({\bf r},{\bf r}')$ and the
self energy $u_\pm({\bf r})$ of the ions, which are the key
equations for weakly coupled electrolytes~\cite{Wang1,Carnie}. Eq.
\ref{eq2.11} has the same form as the PB equation, but now with the
self-energy of the ions appearing in the Boltzmann factor. The
appearance of the self energy in the Boltzmann factor reflects the
nonlinear feedback of the fluctuation effects, an aspect that was
missing in a perturbation expansion.  The self-energy given by Eq.
\ref{eq2.13} is a unified expression that includes the Born energy
of the ion, the interaction between the ion and its ionic
atmosphere, as well as the distortion of the electric field by a
spatially varying dielectric function, the latter taking the form of
image charge interaction near a dielectric discontinuity. In
general, the self energy is spatially varying if there is spatial
inhomogeneity in either the dielectric constant or the ionic
strength. Making use of the variational conditions Eqs. \ref{eq2.11}
and \ref{eq2.12} and evaluating the fluctuation part of the free
energy arising from Gaussian integrals by using the charging method
(as shown in Appendix B), we obtain a simple expression for the
equilibrium grand free energy:
\begin{eqnarray}
W &=&- \int d {\bf r}\left( c_+ + c_- \right) + \frac{1}{2}  \int d {\bf r} \psi  \left( \rho_{ex} - q_+  c_+ + q_- c_-  \right) \nonumber \\
&+&  \int d {\bf r} I({\bf r}) \int_0^1 d \eta  \left[ G ({\bf
r},{\bf r}; \eta) - G ({\bf r},{\bf r})\right] \label{eq2.15}
\end{eqnarray}
where $\eta$ is a ``charging" variable. $G ({\bf r},{\bf r}; \eta)$
is the same-point Green function obtained from solving Eq. \ref{eq2.12} but with the term $I({\bf r})$ replaced with $\eta I({\bf r})$. Note that the free
energy expression Eq. \ref{eq2.15} is finite even in the
point-charge limit.  Although both $G ({\bf r},{\bf r}; \eta)$ and
$G ({\bf r},{\bf r})$ are infinite, their divergent parts exactly
cancel; the remaining difference is finite and accounts for the
leading-order ion-ion correlation effect.  Unlike previous field-theoretical treatments\cite{Netz2,Podgornik2,Dean},
 {\em no microscopic cut-off is needed in our theory}.

\subsection{Weakly Charged Plate}
\label{sec:levelB}

We now specify to the case of a charged plate with dielectric discontinuity in
contact with an electrolyte solution. The fixed external charge
density is then $ \rho_{ex} ({\bf r})= \sigma \delta (z)$.  For concreteness, we take
the surface charge to be positive.  Both $\Gamma$ and
$\varepsilon({\bf r})$ are now step functions: $\Gamma=0$ and
$\varepsilon({\bf r})=\varepsilon_{P}$ for $z<0$; $\Gamma=1$ and
$\varepsilon({\bf r})=\varepsilon_{S}$ for $z>0$. In the solvent
region ($z>0$), Eq. \ref{eq2.11} becomes
\begin{equation}
-\epsilon_{S} \frac{\partial^2 \psi (z)}{\partial z^2}= \lambda_+
q_+ {\rm e}^{ - q_+  \psi -  u_+ } -  \lambda_-  q_- {\rm e}^{q_-
\psi - u_- } \label{eq2.16}
\end{equation}
with the boundary condition $(\partial \psi / \partial z)_{z=0}=-
\sigma /\epsilon_{S}$, which is obtained by integrating Eq.
\ref{eq2.11} between $z=0^-$ and $z=0^+$ and noting that $(\partial
\psi / \partial z) = 0$ for $z<0$. Since the solvent has a uniform
dielectric constant, the Born energy is constant and can be absorbed
into the reference chemical potential. It is then convenient to
single out this constant contribution by rewriting Eq. \ref{eq2.13}
as
\begin{eqnarray}
&&u_{\pm} ({\bf r})   =   \frac{1}{2}   \int d {\bf r}'  d {\bf r}''
h_{\pm} ({\bf r}-{\bf r}') \frac {1}{4 \pi \epsilon_{S} \vert {\bf
r}' - {\bf r}'' \vert  } h_{\pm} ({\bf r}''-{\bf r})\nonumber \\
&&+ \frac{1}{2}   \int d {\bf r}'  d {\bf r}'' h_{\pm} ({\bf r}-{\bf
r}') \left[ G({\bf r}',{\bf r}'')- \frac {1}{4 \pi \epsilon_{S}
\vert {\bf r}' - {\bf r}'' \vert }\right]  \nonumber \\
&& \times h_{\pm} ({\bf r}''-{\bf r})\label{eq2.13a}
\end{eqnarray}

The first term gives a constant Born energy of the ion
$q_{\pm}^2/8\pi \epsilon_{S} a_{\pm}$, with $a_{\pm}$ the Born
radius of the ion\cite{Wang1}. The remaining term is finite in the
point-charge limit.  We can thus take the
$a_{\pm} \to 0$ limit for this term in the final expression, or equivalently and more conveniently by directly
take the point-charge limit in the distribution $h_{\pm} ({\bf r}-{\bf r}')= \delta ({\bf r}-{\bf r}')$.  The
nontrivial and nonsingular part of the self energy $u_{\pm}^{*}$ is then
\begin{equation}
u_{\pm}^{*} ({\bf r})=\frac  {q_{\pm}^2 } {2} \lim_{{\bf r}' \to
{\bf r}} \left[ G({\bf r}, {\bf r}')- \frac{1}{4 \pi \epsilon_{S}
\vert {\bf r} - {\bf r}' \vert}\right] \label{eq2.17}
\end{equation}

To avoid the complexity of solving the equation for the Green
function (\ref{eq2.12}), previous work usually invoked approximate
schemes, e.g., by replacing the spatially varying screening length
by the bulk Debye length
\cite{Monica,onsager,Levinsimulation,Levin1,Levin2} or using a
WKB-like approximation\cite{Carnie,Buff,Levine}. However, the
screening on the image force at the dielectric interface is
inhomogeneous, long-ranged and accumulative, which cannot be
captured fully by these approximate methods.  In this work, we
perform the full numerical solution of the Green function, which
provides the most accurate treatment of the inhomogeneous screening
effect at the dielectric interface.  To solve the Green function in
the planar geometry, it is convenient to work in a cylindrical
coordinate system $(r,z)$. Noting the isotropy and translational
invariance in the directions parallel to the surface, we can perform
a Fourier transform in the parallel direction to write:
\begin{equation}
G(r,z,z')=\frac{1}{2\pi} \int _0 ^\infty kdk J_0(k r) {\hat
G}(k,z,z') \label{eq2.18}
\end{equation}
where $J_0$ is the zero-order Bessel function. ${\hat G}(k,z,z')$ now
satisfies:
\begin{equation}
-\frac{\partial^2 {\hat G}(k,z,z')}{\partial z^2}+\left[\kappa^2 (z)+k^2\right]
{\hat G}(k,z,z')=\frac{1}{\epsilon_S} \delta(z,z') \label{eq2.19}
\end{equation}
for $z>0$, with the boundary condition $\epsilon_S \partial {\hat G}
/
\partial z- k \epsilon_P {\hat
G} =0$ at $z=0$. $\kappa(z)=\left[2 I (z)/\epsilon_{S}\right]^{1/2}$
can be considered the inverse of the local Debye screening length.

Eq. \ref{eq2.19} is solved numerically by using the finite
difference method\cite{Jiaotong,numerical}. The free-space
Green function satisfying $-\partial^2{\hat G}_0/\partial z^2 + k^2 {\hat G}_0
= \delta(z,z')/\epsilon_S$, though analytically solvable, is also solved numerically
along with Eq. \ref{eq2.19} to ensure consistent numerical accuracy in removing the singularity of the
same-point Green function.  The nondivergent part
of the self energy is then:
\begin{equation}
u_{\pm}^{*} (z)=\frac{q_{\pm}^2}{4\pi} \int _0 ^\infty  \left[ {\hat
G}(k,z,z) -  {\hat G}_0(k,z,z)\right] kdk \label{eq2.20}
\end{equation}

Far away from the plate surface ($z \to \infty$), the ion
concentration approaches the bulk value $c^b_{\pm}$, and from Eq.
\ref{eq2.14} (where we set $\psi_b=0$, or equivalently absorbing a constant $\psi_b$ into the definition of the fugacity),  the fugacity of the ions is given by
$\lambda_\pm=c^b_{\pm} \exp\left[ -q^2_\pm \kappa_b/(8\pi
\epsilon_{S}) \right]$ where $\kappa_b$ is the inverse screening
length in the bulk.  Note that this relationship automatically takes into account the Debye-H\"uckel correction to fugacity due to ion-ion correlations.

The theory presented above is derived explicitly with added salt.  However, application to the
counterion-only system is straightforward through an ensemble
transformation\cite{Netz2}.

\section{Numerical Results and Discussions}
\label{sec:level3}

In
this section, we apply the theory presented in the last section to an electrolyte solution in
contact with a weakly charged plate.  We first examine the counterion-only system to highlight the depletion boundary layer issue and then study the consequences of the depletion boundary layer on the structure and thermodynamic properties of the electric double layer with added salts.

\subsection{Counterion-only Case}

For the counterion-only system, the PB theory admits an analytical
solution for the counterion distribution: $c(z)=1/\left[2 \pi l_B
q^2 (z^2+l_{GC}^2)\right]$, which is characterized by a single
length scale, the Gouy-Chapman length.  The counterion concentration
profile is shown in Fig. 1 as the dashed line, which decays
monotonically.    In contrast, when there is dielectric
discontinuity, our theory predicts a qualitatively different
behavior.  The presence of the depletion boundary layer inside the
Gouy-Chapman length is obvious, and is consistent with results from
Monte Carlo simulation\cite{Netz,Levinsimulation}. Within the
depletion boundary layer ($z<d \sim l_B \Xi^{-1/2}$), image charge
repulsion is dominant and ions are excluded from the plate surface.
In the point-charge model, the self energy diverges to infinity at
the plate surface; thus the ion concentration vanishes at $z=0$. The
vanishing of the ion concentration obviously contracts the PB
prediction but is also incapable of being captured by any
perturbation corrections around the PB
limit\cite{Ninham1,Ninham2,Netz2,Kanduc,Podgornik,Podgornik2,Podgornikrev}.
Simply put, these perturbative approaches fail to satisfy the
boundary condition for the ion concentration at $z=0$, as is typical
with boundary layer problems. Beyond the depletion boundary layer
($z>d$), surface charge attraction prevails and the ion
concentration approaches the PB profile sufficiently far away from
the surface.

\begin{figure}[hbt]
\centering
\includegraphics[width=0.45\textwidth]{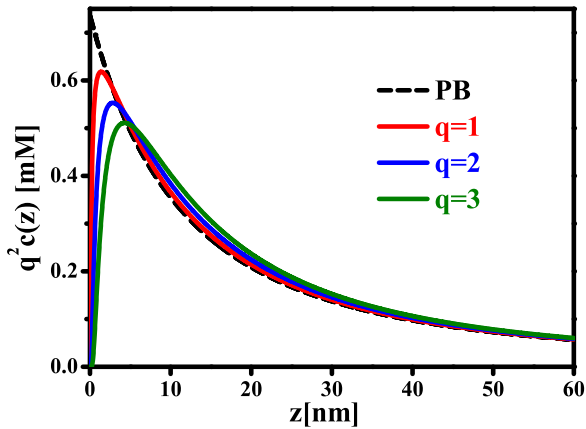}
\caption{Ion concentration for the counterion-only system for
different ion valency $q$. $\varepsilon_S=80$ and
$\varepsilon_P=2.5$. $\sigma=1/(100q) (e/nm^2)$. The Gouy-Chapman
length is kept constant ($l_{GC}=22.7nm$) for counterions of
different valencies. The coupling parameter $\Xi$ is $0.031 q^2$.
 \label {2}}
\end{figure}

The PB theory predicts a universal profile $q^2 c(z)$ for
counterions of different valencies when the Gouy-Chapman length is
kept fixed.  However, from our scaling analysis in the Introduction,
the depletion boundary layer should increase linearly with valency;
see Eq. \ref{eq1.1}.  This prediction is borne out by our numerical
result as shown in Fig. 1.  Therefore, the boundary layer problem
becomes more severe for ions of high valency. The scaling of the
boundary thickness with the coupling parameter predicted from Eq.
\ref{eq1.2} is also confirmed by our numerical results (data not
shown).

\subsection{With Symmetric Salt}

When there are added salt ions in the solution,  the image force
affects the distribution of both the counterions and coions. The PB
theory predicts that the double layer structure is characterized by
the Debye screening length $\kappa^{-1}$ under the condition that
$\kappa^{-1} \ll l_{GC}$, with a monotonically decreasing counterion
and monotonically increasing coion distribution.  In contrast, both
the counterion and coion concentration must vanish at the surface,
but their approach to the bulk concentration is different: the
coions increases monotonically, while the counterions goes through
an overshoot.  Furthermore, we find two regimes depending on the
relative width of the screening length and the boundary layer
thickness, which itself is in turn affected by the screening.  At
low salt concentration, $\kappa^{-1} \gg d$ and ion depletion is
confined in a boundary layer very close to the plate surface; both
the ion distribution and electrostatic potential approach the
profile predicted by PB beyond the boundary layer.  As the salt
concentration increases, the width of the depletion boundary layer
becomes comparable to the screening length and the two length scales
remain comparable thereafter; the image charge interaction then
affect the entire range of the double layer. In Figure 2 we show the
ion distribution of a $0.1M$ 1:1 electrolyte calculated by our
theory. The contrast with the PB result is quite striking.

\begin{figure}[hbt]
\centering
\includegraphics[width=0.45\textwidth]{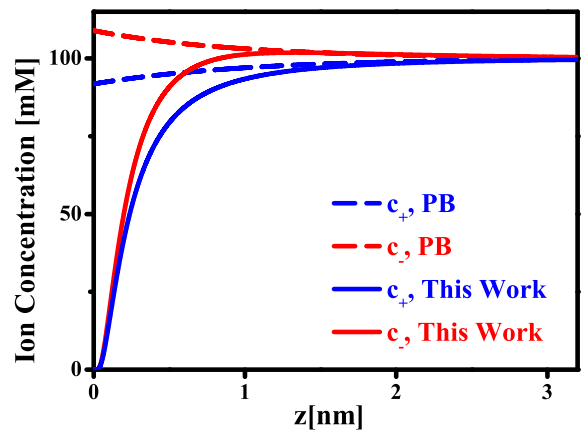}
\caption{Ion concentration profile for a 1:1 electrolyte solution with
$c^b=0.1M$. $\varepsilon_S=80$,
$\varepsilon_P=2.5$ and $\sigma=1e/100nm^2$.
 \label {5}}
\end{figure}

\begin{figure}[b]
\centering
\includegraphics[width=0.45\textwidth]{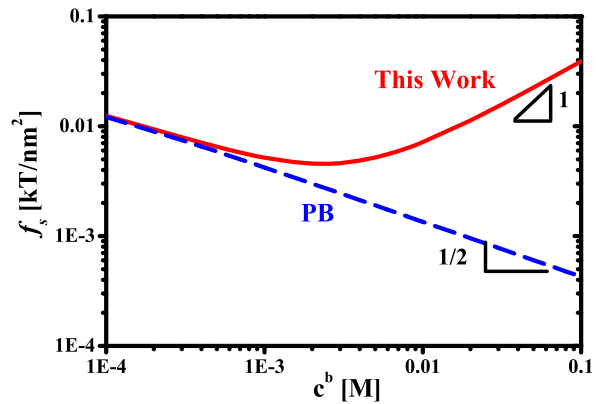}
\caption{The surface energy $f_s$ as a function of the salt
concentration for a 1:1 electrolyte solution. $\varepsilon_S=80$,
$\varepsilon_P=2.5$ and $\sigma=1e/100nm^2$.
 \label {5}}
\end{figure}

The change in the double layer structure will affect a wealth of
interfacial properties.  As an example, we show in Figure 3 the
surface excess free energy $f_s= \int_0 ^{\infty} (w-w^{b})dz$
(where $w$ is the grand free energy density and $w^{b}$ is its bulk
value) as a function of the salt concentration. The PB theory
predicts a monotonic decrease of $f_s$ that scales approximately
with $(c^b)^{-1/2}$, which arises from the electric field
contribution in the free energy due to the surface
charge\cite{Onuki,Wang4}. With the inclusion of image charge
interaction, our theory shows that $f_s$ changes nonmonotonically.
At low salt concentration ($c^b < 10^{-3}M$), $f_s$ calculated by
our theory follows closely the PB result; this is because the region
affected by the image charge repulsion is relatively narrow compared
to the screening length, giving a relatively small contribution to
the surface excess energy when integrated over the entire solution.
As the salt concentration increases ($c^b
> 10^{-2}M$), our theory predicts a sign change in the slope of
$f_s$ vs. $c^b$: $f_s$ increases with increasing $c^b$, {\em
opposite to} the PB result.  In this concentration regime, the width
of the depletion boundary layer is comparable to the Debye screening
length, and the entire double layer region is affected by the image
charge interaction as shown in Figure 2.  The increase in $f_s$ is
now largely due to the depletion (i.e., negative adsorption) of
mobile ions.  The slope of $\log(f_s)$ vs $\log(c^b)$ is less than
$1$ because of the increased screening of the image force as the
salt concentration increases. The sign change of $\partial f_s
/\partial c^b$ corresponds to the crossover in the length scale
relationship from $\kappa^{-1} \gg d$ to $\kappa^{-1} \approx d$. As
the excess surface energy determines the spreading of a liquid drop
on a solid surface, this result implies a qualitatively different
behavior for the spreading of a drop of electrolyte solution than
that predicted by the PB theory.  We also note that the nonmonotonic
behavior discussed here shares the same physics as the Jones-Ray
effect\cite{Onuki,JonesRay,Bier,Wang4} for the interfacial tension
observed at the water/air and water/oil interfaces.

\subsection{With Asymmetric Salt}

The effects of image charge become more complex if the salt ions are
of unequal valency.  Because of the quadratic dependence of the
image force on the valency, the higher-valent ions are pushed
further away from the surface, necessitating a compensation by the
lower-valent ions in the space in between. The difference in the
image force between the counterions and the coions induces
additional charge separation and hence electric field within the
depletion boundary layer. The induced net charge within the boundary
layer alters the effective surface charge, which can affect the
double layer structure outside the boundary layer.  For the case
where the coions are of higher valency than the counterions, the
induced electric field due to unequal ion depletion counteracts the
field generated by the surface charge.  With the increase of the
salt concentration, the induced field can exceed that generated by
the bare surface charge, leading to a sign change in the effective
surface charge known as charge inversion.  The double layer
structure becomes qualitatively different from that predicted by the
PB theory as shown in Figure 4: the electrostatic potential is of
the opposite sign to the PB result.  Excess counterions accumulate
in the depletion boundary layer, overcharging the plate surface,
while the coions are enriched outside the boundary layer, serving to
screen the inverted surface charge.  In this case, the PB theory
qualitatively fails to describe the entire double layer structure.

\begin{figure}[th]
\centering
\includegraphics[width=0.45\textwidth]{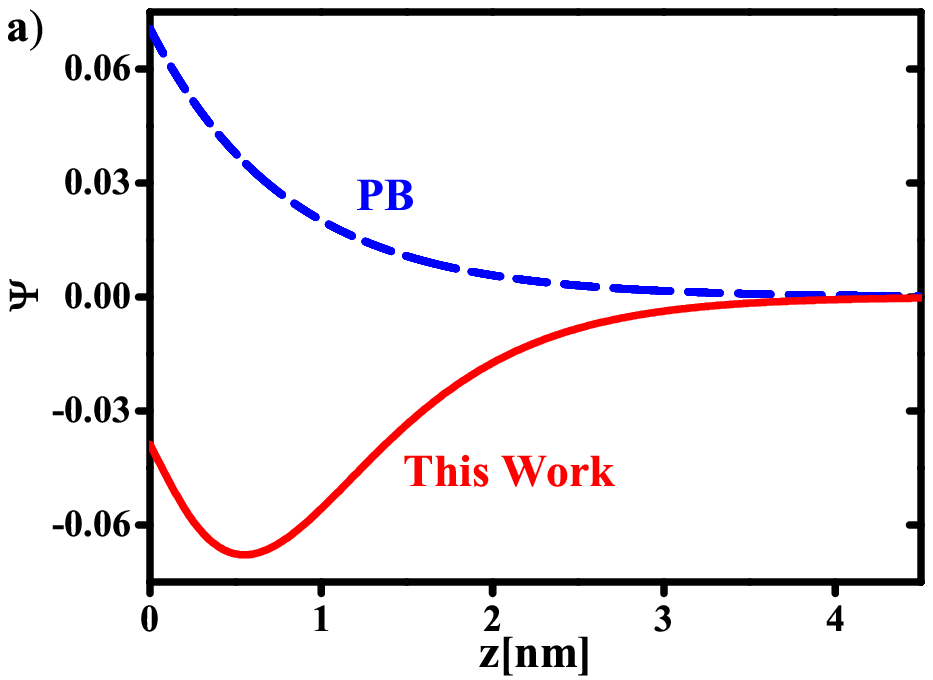}
\includegraphics[width=0.45\textwidth]{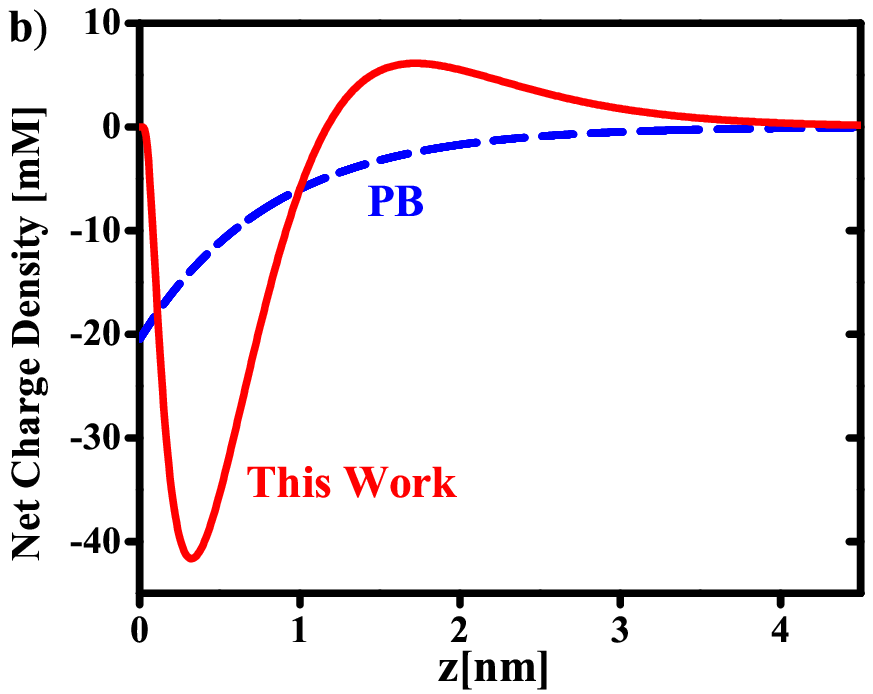}
\caption{Charge inversion for a 0.05M 2:1 electrolyte solution near
a positively charged plate. (a) Dimensionless electrostatic
potential and (b) net charge density ($q_+ c_+ -q_- c_-$).
$\varepsilon_S=80$, $\varepsilon_P=2.5$ and $\sigma=1e/100nm^2$.
 \label {5}}
\end{figure}

\subsection{Uncharged Surface: Image Charge vs. Correlation Effect}

The case of an electrolyte solution next to an uncharged surface
($\sigma=0$) reduces to the problem treated by Wagner, Onsager and
Samaras.  The self energy due to image charge repulsion appears in
the Boltzmann factor and is responsible for the depletion layer in
the ion distribution near the surface as shown in Figure 5. Note,
however, in the original WOS theory as well as in subsequent
treatments \cite{Monica,onsager,Levinsimulation,Levin1,Levin2}, the
image charge term was added to the Boltzmann factor {\em ad hoc}
based on physical intuition, whereas in our theory, its appearance
is the result of systematic derivation.  Therefore, our theory not
only recovers the WOS theory (upon making additional approximations,
e.g., by using the constant bulk screening length for the image
force potential) but also provides the means for systematically
improving the WOS theory. First, our theory captures the anisotropic
screening cloud around an ion near the interface due to the
spatially varying ion concentration near the surface. The
inhomogeneous ionic cloud in the depletion layer and its effect on
the screening of the test ion are treated self-consistently in our
theory, whereas this inhomogeneous screening is missing in the WOS
theory.  Second, by including the mean electrostatic potential
generated by charge separation, our theory can describe salt
solutions with unequal valency such as the case of 2:1 electrolyte
shown in Figure 5b.  Finally, our theory provides a more accurate
expression for the excess free energy by properly accounting for the
inhomogeneous screening effect and the fluctuation contribution to
the free energy.  Thus, we expect our theory to be able to better
predict the surface tension of electrolyte solutions in comparison
to the WOS theory, especially at higher salt concentrations (where
accurate treatment of the screening becomes more important).

The inhomogeneous screening results in a correlation effect that can lead to ion depletion near the surface\cite{Monica}: an ion interacts
more favorably with its full ionic atmosphere far away from the
surface than in the vicinity of the surface.  This correlation
effect is stronger for multivalent ions, which pushes them further
away from the interface than the monovalent ions. The
correlation-induced ion depletion near the surface can take place
both with and without the dielectric contrast, and is well captured by our theory, as shown in Figure 5.
While the ion depletion without the dielectric contrast
is induced by the correlation alone, the ion depletion in the presence of the dielectric contrast is due to both the correlation
effect and the image charge effect, which enhance each other.  As a result, both ion
depletion, as well as charge separation in the case of 2:1
electrolyte, are more pronounced in the presence of image charge than due to correlation alone.

Ion depletion due to correlation alone is most noticeable when the
surface is uncharged.  When the surface is charged, the surface
attraction for the counterions will dominate over such correlation
effect in the absence of image charge repulsion.  In contrast,
depletion due to image charge repulsion persists for both the
counterions and coions even when the surface is charged.

\begin{figure}[ht]
\centering
\includegraphics[width=0.45\textwidth]{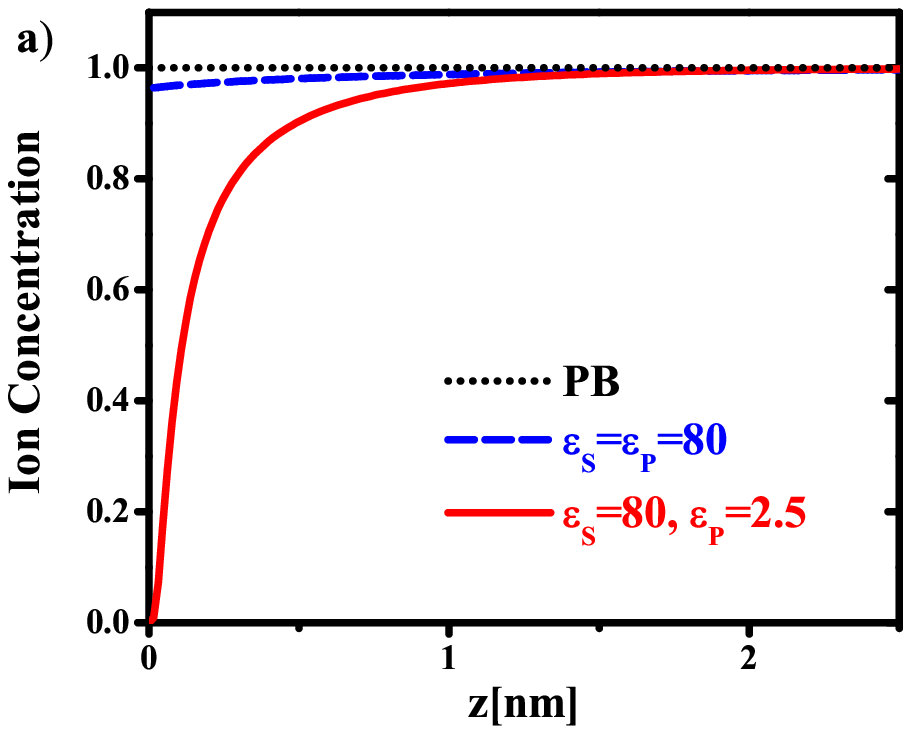}
\includegraphics[width=0.45\textwidth]{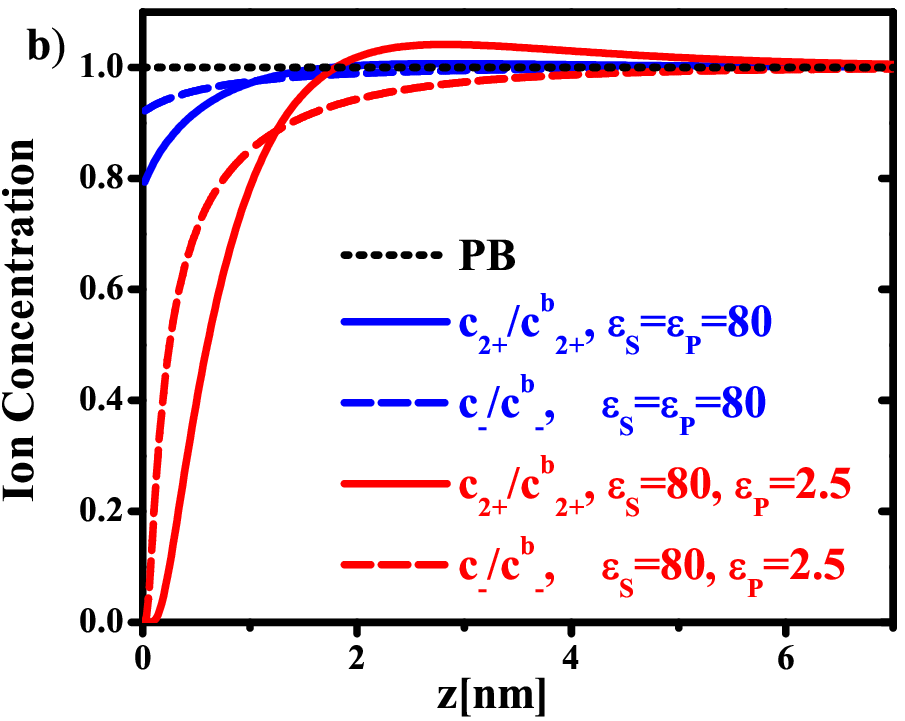}
\caption{(Color online) Ion concentration (scaled by the bulk ion
concentration $c_{\pm}^{b}$) for (a) 0.01M 1:1 electrolyte solution
($c_{+}=c_{-}$) and (b) 0.01M 2:1 electrolyte solution near an
uncharged interface ($\sigma=0$) with dielectric contrast
($\varepsilon_S=80$, $\varepsilon_P=2.5$) in comparison with the
case without dielectric contrast ($\varepsilon_S=\varepsilon_P=80$).
Profiles calculated by our theory are shown by colored lines; results from the PB theory are given as black dot lines.
 \label {5}}
\end{figure}

\section{Conclusions}
\label{sec:level3}

In this work, we have shown that the image charge repulsion creates
a depletion boundary layer near a dielectric surface, which cannot
be captured by a regular perturbation method.  Using a
nonperturbative approach based on a variational functional
formulation, we find that the self energy of the ion, which includes
contributions from both the image charge interaction and the
interionic correlation, appears explicitly in the Boltzmann factor
for the ion distribution, resulting in a self-energy modified
Poisson-Boltzmann equation as the appropriate theory for describing
the {\em physical} weak-coupling condition.  This image-charge self
energy is not diminished by reducing the surface or the ionic
strength in the solution;  in the presence of a significant
dielectric discontinuity, there is no limiting condition for which
the PB theory is valid.  For zero surface charge, our theory reduces
to the WOS theory upon further approximations.  Thus,
our theory provides both the justification for the WOS theory and
means for systematically improving the WOS theory, for example, by
including the mean electrostatic potential generated by the charge
separation in salt solutions with unequal valency or other
asymmetries between the cation and anions, such as different size
and polarizability\cite{Levin1}.

The weak-coupling condition in the presence of dielectric
discontinuity covers many soft-matter and biophysical systems.  Many
phenomena, such as the surface tension of electrolyte
solutions\cite{Surfacetension1,Surfacetension2}, salt effects on
bubble coalescence\cite{Bubble}, and the ion conductivity in
artificial and biological
ion-channels\cite{Channel1,Channel2,Channel3}, cannot be explained,
even qualitatively, by the PB theory. The presence of the image
charge interaction results in a very different picture of the
electrical double layer from that provided by the PB theory, and can
give rise to such phenomena as like-charge attraction and charge
inversion even in the weak-coupling condition\cite{Wang3}; these
phenomena have usually been associated with the strong-coupling
condition. The PB theory has played a foundational role in colloidal
and interfacial sciences: the DLVO theory,  interpretation of the
zeta potential, experimental determination of the surface charge and
the Hamaker constant, are all based on the PB
theory\cite{Israelachvili}.  With the inclusion of the image charge
interaction, some of the well known and accepted results will have
to be reexamined.

\begin{acknowledgments}
Acknowledgment is made to the donors of the American Chemical
Society Petroleum Research Fund for partial support of this
research.
\end{acknowledgments}

\appendix
\section{Derivation of the key equations in Section II.A}

We define $\chi \equiv \phi+ i \psi$ as the fluctuation
part of the field $\phi$, which is a Gaussian variable by our ansatz.  The
variational grand free energy can be approximated by the r.h.s of
Eq. \ref{eq2.8} as
\begin{eqnarray}
&W& = -\ln \Omega_{ref}+ \langle L\left[ \phi \right] - L_{ref} \left[ \phi \right] \rangle = - \frac{1}{2} \ln  \left(\frac{\det {\bf G}}{\det {\bf G}_0} \right) \nonumber \\
&-&  \frac{1}{2}   \int d {\bf r} d {\bf r}' \{ \delta ({\bf r}' - {\bf r} ) \left[ \epsilon (\nabla \psi)^2 - \epsilon \langle (\nabla \chi)^2 \rangle \right] + G^{-1} ({\bf r},{\bf r}')        \nonumber \\
&\times& \langle \chi({\bf r}) \chi({\bf r}') \rangle  \} + \int d
{\bf r} [ \rho_{ex} \psi - \Gamma \lambda_+ {\rm e}^{- q_+ \psi }
\langle {\rm  e}^{- i {\hat h}_+ \chi }
\rangle  \nonumber\\
&+&  \Gamma \lambda_- {\rm  e}^{ q_- \psi } \langle {\rm  e}^{ i
{\hat h}_- \chi } \rangle   ] \label{eqA1}
\end{eqnarray}
The averages in Eq. \ref{eqA1} can be evaluated exactly because the
distribution of $\chi$ is Gaussian. Noting that
\begin{equation}
\langle \chi({\bf r}) \chi({\bf r}') \rangle =G ({\bf r}, {\bf r}')
\label{eqA2}
\end{equation}
we have
\begin{eqnarray}
\int d {\bf r} d {\bf r}' \delta ({\bf r}- {\bf r}') \epsilon
\langle (\nabla \chi)^2 \rangle \nonumber\\
= \int d {\bf r} d {\bf r}' \nabla _{\bf r} \cdot \left[\epsilon
({\bf r}) \nabla _{\bf r}' \delta ({\bf r}-{\bf r}')  \right] G({\bf
r},{\bf r}') \label{eqA3}
\end{eqnarray}
and
\begin{eqnarray}
\langle {\rm  e}^{\mp i  {\hat h}_{\pm} \chi } \rangle =\exp [
&-&\frac{1}{2} \int d {\bf r}'  d {\bf r}''
h_{\pm} ({\bf r}-{\bf r}') G({\bf r}',{\bf r}'')\nonumber\\
&\times& h_{\pm} ({\bf r}''-{\bf r}) ] \label{eqA4}
\end{eqnarray}
Substituting Eqs. \ref{eqA2}-\ref{eqA4} into Eq. \ref{eqA1}, we
obtain the variational form of the grand free energy as
\begin{eqnarray}
&W& = -\frac{1}{2} \ln  \left(\frac{\det {\bf G}}{\det {\bf G}_0} \right) - \frac{1}{2} \int d{\bf r}  \epsilon (\nabla \psi)^2  \nonumber \\
&-& \frac{1}{2}  \int d {\bf r} d {\bf r}'\left[  G^{-1} ({\bf
r},{\bf r}') -G_0^{-1} ({\bf r},{\bf r}') \right] G ({\bf r},{\bf
r}')\nonumber \\
&+& \int d {\bf r} \left(  \rho_{ex} \psi -\Gamma \lambda_+ {\rm
e}^{- q_+ \psi -u_+} - \Gamma \lambda_- {\rm e}^{ q_- \psi - u_-}
\right) \label{eqA5}
\end{eqnarray}
where $u_{\pm}$ is the self energy of the ions given by Eq.
\ref{eq2.13}. Minimizing Eq. \ref{eqA5} with respect to $\psi$ and
$G$ gives rise to Eq. \ref{eq2.11} and Eq. \ref{eq2.12} in the main
text.

\section{Simplification of the fluctuation contribution in the free energy}

Making use of Eq. \ref{eq2.11} and Eq. \ref{eq2.12}, Eq. \ref{eqA5}
can be simplified as
\begin{eqnarray}
W &=&- \int d {\bf r}\left[ c_+({\bf r}) + c_-({\bf r}) \right] \nonumber \\
&+& \frac{1}{2}  \int d {\bf r} \psi ({\bf r}) \left[ \rho_{ex} ({\bf
r})- q_+  c_+({\bf r}) + q_- c_-({\bf r})  \right] \nonumber \\
&-&\frac{1}{2} \ln  \left(\frac{\det {\bf G}}{\det {\bf G}_0}
\right) - \int d {\bf r} I({\bf r}) G ({\bf r},{\bf r}) \label{eqB1}
\end{eqnarray}
The last two terms in Eq. \ref{eqB1} are due to the fluctuation
contribution. We note that the Green function equation (Eq.
\ref{eq2.12}) can be written in the matrix form as
\begin{equation}
{\bf G}_0^{-1} {\bf G} + 2 I({\bf r})  {\bf G}= {\bf I} \label{eqB2}
\end{equation}
where  ${\bf I}$ is the identity matrix (not to be confused with the
local ionic strength $I$). Right multiplication of the above
equation by ${\bf G}^{-1}$, we obtain
\begin{equation}
{\bf G}^{-1} = {\bf G}_0^{-1} + 2 I({\bf r}) {\bf I} \label{eqB3}
\end{equation}
Note also that
\begin{eqnarray}
 \ln  \left(\frac{\det {\bf G}}{\det {\bf G}}_0 \right) =  \ln  \det {\bf G} - \ln \det {\bf G}_0  \nonumber \\
= \int d {\bf r}\int d{\bf r}' \int_{G_0^{-1}}^{G^{-1}}\frac{\delta
\ln \det {\bf G}}{\delta G^{-1}({\bf r},{\bf r}')} \delta
G^{-1}({\bf r},{\bf r}') \label{eqB4}
\end{eqnarray}
The innermost integral is a functional integration over $G^{-1}$
from $G_0^{-1}$ to $G^{-1}$. Since $\ln \det {\bf G}$ is the result
of a Gaussian functional integral, we have
\begin{equation}
 \frac{\delta \ln \det {\bf G}}{\delta G^{-1}({\bf r},{\bf
r}')}=- G({\bf r},{\bf r}') \label{eqB5}
\end{equation}
Therefore,
\begin{equation}
\ln  \left(\frac{\det {\bf G}}{\det {\bf G}_0} \right)=-\int d {\bf
r}\int d{\bf r}' \int_{G_0^{-1}}^{G^{-1}} G({\bf r},{\bf r}') \delta
G^{-1}({\bf r},{\bf r}') \label{eqB6}
\end{equation}
As the integration goes from  $G_0^{-1}$ to $G^{-1}$, the integrand
changes from $G_0$ to $G$. From Eq. \ref{eqB3}, a convenient path
for integrating Eq. \ref{eqB6} is to introduce a continuous
``charging" variable $\eta$ that goes from 0 to 1 multiplying the $2
I$ term in Eq. \ref{eqB3}, while keeping the density profile fixed.
Obviously the Green  function is $G_0$ for $\eta=0$ and is $G$ for
$\eta=1$. For any intermediate value, we denote the Green function
as $G({\bf r},{\bf r}'; \eta)$. Using Eq. \ref{eqB3}, the above
integral becomes,
\begin{eqnarray}
&\ln&  \left(\frac{\det {\bf G}}{\det {\bf G}_0} \right)=-2 \int d {\bf r}\int d{\bf r}' \int_0^1 I({\bf r}) \delta ({\bf r}-{\bf r}')  \nonumber  \\
&\times& G({\bf r},{\bf r}';\eta) d \eta =- 2 \int d {\bf r} I({\bf
r}) \int_0^1  G({\bf r},{\bf r}; \eta) d \eta \label{eqB7}
\end{eqnarray}
where $G({\bf r},{\bf r}; \eta)$ is to be understood as the limit
$G({\bf r},{\bf r}; \eta) = \lim_{{\bf r}'\rightarrow {\bf r}}
G({\bf r},{\bf r}';\eta)$, and the Green function $G({\bf r},{\bf
r}';\eta)$ is the solution of
\begin{equation}
- \nabla \cdot \left[ \epsilon \nabla G({\bf r},{\bf r}') \right] +
2 \eta I({\bf r}) G({\bf r},{\bf r}')=  \delta ({\bf r}-{\bf r}')
\label{eqB8}
\end{equation}
With Eq. \ref{eqB7}, the fluctuation contribution to the free energy
is
\begin{eqnarray}
 \frac{1}{2} \ln  \left(\frac{\det {\bf G}}{\det {\bf G}_0} \right) +    \int d {\bf r} I({\bf r}) G ({\bf r},{\bf
 r}) \nonumber  \\
= -\int d {\bf r} I({\bf r}) \int_0^1 d \eta  \left[ G ({\bf r},{\bf
r}; \eta) - G ({\bf r},{\bf r})\right] \label{eqB9}
\end{eqnarray}
which is finite even in the point-charge limit.

\end{document}